\documentclass{epl}
\usepackage{amssymb}
\usepackage{amsmath}
\usepackage{amscd}

\title{Optimal control of dissipation
for the example\\
of the spin--boson model}

\author{H. Jirari\thanks{hamza.jirari@uni-graz.at}
and W.~P\"{o}tz\thanks{walter.poetz@uni-graz.at}}

\institute{ Institut f\"{u}r Physik, Theory Division
Karl-Franzens-Universit\"{a}t Graz,\\
Universit\"{a}tplatz 5, 8010 Graz, Austria}


\newcommand{\be}{\begin{equation}}
\newcommand{\ee}{\end{equation}}
\newcommand{\bea}{\begin{eqnarray}}
\newcommand{\eea}{\end{eqnarray}}

\newcommand*{\bs}{\begin{split}}
\newcommand*{\bes}{\begin{equation}\begin{split}}
\newcommand*{\ees}{\end{split} \end{equation}}
 \reversemarginpar 
\begin{document}
\maketitle

\begin{abstract}
The interaction of a quantum system with a bath, usually 
referred to as dissipation,  can be controlled if one 
can establish quantum interference between the system--bath 
interaction and a coupling of the system to an external control field. 
This is demonstrated for the example of the spin-boson 
model in the strong coupling limit for the  system--bath interaction.   
It is shown that driving and trapping of the spin system leads 
to an optimum control problem which is nonlinear in the 
external control field. Using an indirect optimization strategy 
introducing a Lagrange-type adjoint
state, we show that the spin system can be trapped 
in otherwise unstable quantum states and that 
it can be driven from a given initial state to a 
specified target state with high fidelity.
\end{abstract}

\section{Introduction}
 
Over a last a few years, a number of interesting schemes have been 
proposed to eliminate the undesirable effects of decoherence
in open quantum systems, including decoherence free subspaces~\cite{Zanardi_1,Lidar}, 
quantum error correction codes~\cite{Nielson,Preskill,Knill}, quantum feedback
~\cite{Wang} and mechanisms based on unitary "bang-bang" 
pulses and their generalization, quantum dynamical decoupling
~\cite{Viola,Vitali}.
The key ingredient of dynamical decoupling
is the continuous disturbance of the system, which suppresses the system-environment
interaction. It has been shown that, in the bang-bang control schemes,
the decoherence of the system is effectively suppressed if the pulse
rate is much higher than the decoherence rate due to the system-environment 
interaction~\cite{Viola}. 

The staring point of the decoupling techniques is the observation that even
though one does not have access to the large number of uncontrollable degrees
of freedom of the environment, it is still possible to interfere with its dynamics
by inducing interactions into the sub--system which drive it so fast that the environment
cannot follow~\cite{Vitali}. Alternatively, 
if one can establish a suitable coupling to the system   
by means of an external control, one can establish quantum interference 
with the system--bath.  In a simple minded model for a dissipative quantum system, 
where the interference between the system--bath and system--control 
interaction is ignored or is irrelevant only limited control 
can be achieved~\cite{Jirari}.
The situation changes dramatically when 
interference between the system--environment and system--control 
interaction can be used to control the effective system--environment
coupling~\cite{Viola,Vitali,Hartmann,Kohler,Allahverdyan,CCP,Xu,
Kocharovskaya,Schirr,Morillon,Dakhnovskii,Grifoni}.  
The degree and nature of quantum interference 
constructive or destructive can be controlled by adjustment of the control field, known as coherent control.

In this paper we apply the concept of coherent control to steer a dissipative quantum system 
to the spin boson model, in which a quantum two-level system (qubit)
is modelled by a spin, the environmental heat bath by quantum oscillators 
and the spin subjected to external control field is coupled
to each bath oscillator independently~\cite{Grifoni,WEISS,Legget}.
Achieving decoherence control for this model is formulated using optimal control 
which is mathematically a problem of functional optimizations
under constraints in form of differential equations~\cite{Jirari,ARTHURE}. 
The two-level system coupled to a bath provides an adequate model
of such diverse phenomena as electron transfer reaction~\cite{Morillon},
electron--phonon interaction in point defects~\cite{Kuhn} and quantum dots~\cite{Hohenester}, 
interacting many--body systems~\cite{Zhang}, magnetic
molecules\cite{Luban} and bath assisted cooling of spins~\cite{Allahverdyan}.

\section{Bloch-Redfield formalism}
Consider a physical system $S$ embedded in a dissipative environment $B$, also 
referred to as the heat bath, and interacting with a time-dependent classical 
external field {\it i.e.}, the ``control".
The total Hamiltonian
$
H_{\rm tot} =H_S(t)+ H_B+ H_{\rm int}
$
is the sum of the system Hamiltonian $H_S(t)$, the bath Hamiltonian $H_B$ 
and their interaction $H_{\rm int}$, which is responsible for decoherence. 
Note that the operator $H_S(t)$ contains a time-dependent external field to control
the quantum evolution of the system. We suppose that the system-environment interaction
Hamiltonian is bilinear 
$
H_{\rm int}= \sum_{\alpha}\, A_\alpha\otimes B_\alpha
$
where $A_\alpha$ and $B_\alpha$ are Hermitian operators 
of the system and the environment, respectively.

The basic assumptions underlying the derivation of the 
equation of motion for the reduced density matrix
$
\rho(t)= \mbox{Tr}_B\left\{\rho_{\rm tot}(t)\right\},
$
are that
(i) the initial factorization ansatz; we assume that
at time $t=0$ the bath $B$ is in thermal equilibrium
and uncorrelated with the system $S$ 
($\rho_{\rm tot}(0)=\rho(0)\otimes\rho_B$, Feynman-Vernon approximation),
(ii) weak system-bath interaction limit in which the second-order
perturbation theory is applicable 
($\rho_{\rm tot}(t)=\rho(t)\otimes\rho_B+ \mathcal{O}(H_{\rm int})$, Born approximation)
(iii) the relaxation time $\tau_B$ of the heat bath is much shorter than the time
scale $\tau_R$ over which the state of the system varies appreciably 
($\tau_B \ll \tau_R$, justifying the Markov approximation). From the Liouville-von Neumann equation
$i\hbar\dot\rho_{\rm tot}=\left\lbrack H_{\rm tot},\rho_{\rm tot}\right\rbrack$
for the total density operator and after performing the above
mentioned approximations, one obtains the master equation for the reduced density matrix
in Bloch-Redfield form 
\be
\dot{\rho}_{ij}=
-\frac{i}{\hbar}\sum_{kl}\left({H_S}_{ik}(t)\delta_{lj}-\delta_{ik}{H_S}_{lj}(t)\right){\rho}_{kl}
-\sum_{kl}{\mathcal{R}}_{ijkl}(t)\,{\rho}_{kl}\,.
\ee
where the first term represents the unitary part of the dynamics
generated by the system Hamiltonian $H_S(t)$ and the second term 
accounts for dissipative effects of the coupling of the system to the environment.
The Redfield relaxation tensor $R_{ijkl}(t)$ is given by ~\cite{Blum}
\bea
&& R_{ijkl}(t)=\delta_{lj}\sum_r\,\Gamma_{irrk}^+(t)
+\delta_{ik}\sum_r\,\Gamma_{lrrj}^-(t)
-\Gamma_{ljik}^+(t) -\Gamma_{ljik}^-(t)\, ,
\eea
where the time-dependent rates $\Gamma_{ijkl}^{\pm}(t)$ are evaluated
through the following expressions:
\bea
\label{eq:Gamma+_BIS}
&&\Gamma_{lj,ik}^+(t)=\frac{1}{\hbar^2}\int_0^{t}d\tau
\sum_{\alpha,\beta}\left\langle B_\alpha(\tau)B_\beta(0)\right\rangle_B
A_{\alpha,{lj}}\sum_{m,n} U_{S,{im}}(t,t-\tau)A_{\beta,{mn}} U_{S,{kn}}^{\ast}(t,t-\tau)\, ,\\
\label{eq:Gamma-_BIS}
&&\Gamma_{lj,ik}^-(t)=\frac{1}{\hbar^2}\int_0^{t}d\tau
\sum_{\alpha,\beta}\left\langle B_\beta(0)B_\alpha(\tau)\right\rangle_B 
\sum_{m,n}U_{S,{lm}}(t,t-\tau)A_{\beta,{mn}}U_{S,{jn}}^{\ast}(t,t-\tau)\,A_{\alpha,{ik}}\, .
\eea
Here
$
U_S(t,t')=\mbox{T}
\left\{
\exp\left\lbrack -\frac{i}{\hbar}\int_{t'}^t\,d\tau\,H_S(\tau)\right\rbrack
\right\}
$ is the propagator of the coherent system dynamics while 
$
\langle B_\alpha(t)B_\beta(0)\rangle =\mbox{Tr}_B\left\{B(t)B(0)\rho_B\right\}
$
is the environment correlation function with
$\rho_B = \exp(-\beta H_B)/Z$, the canonical ensemble of the 
bath at the inverse temperature $\beta=1/k_BT$.

Because of the applied control field, the transition 
rates defined by  
$
W_{jl}(t) = \Gamma_{lj,jl}^+(t) + \Gamma_{lj,jl}^-(t)\,,
$
in the secular approximation which we suppose also valid in the driven case,
become time dependent~\cite{Kocharovskaya,Schirr}.
The field influence on both the unitary and dissipative contributions to the 
time evolution of the physical system makes possible an external control of dissipation.
In particular, the correlation between the control field and the dissipation 
leads to the destruction of the detailed balance
$
\lim_{t \to \infty}{W_{ij}(t)}/{W_{ji}(t)}\not={\exp(-\beta E_i)}/{\exp(-\beta E_j)}\,
$
where $E_i$ are the energy eigenvalues of the undriven physical system.
So, the steady state can be far from equilibrium in the driven case.
The influence of the control field on the relaxation tensor via 
$U_S(t,t')$ is a direct consequence of 
quantum interference between the system-bath interaction
and the coupling of the system to the external field.
\begin{figure}
\onefigure[width=6cm,angle=-90]{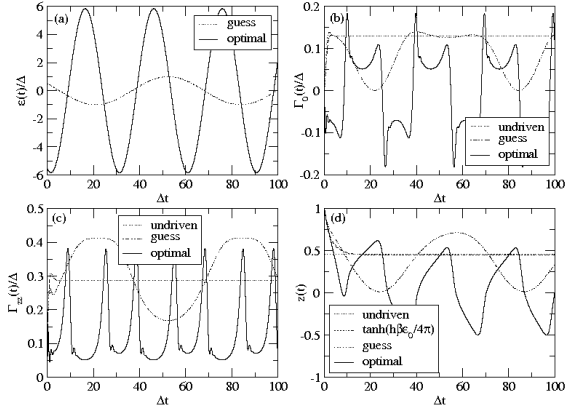}
\caption{Population transfer from the ground state $z(0)=1$ to the 
maximally mixed state $z(t_f)=0$.
(a) shows the control field vs. time.
(b) and (c), respectively, show  $\Gamma_{zz}$ and $\Gamma_{0}$ vs. time.
(d) shows the relative population $z=\left\langle\sigma_z\right\rangle$ 
vs. time. $w_f=1$, $w_r=0$, $\alpha=0.2$, $\epsilon_0=\Delta$,
$\omega_c=20\Delta$, $\beta=1/\hbar\Delta$ 
and $t_f= 100/\Delta$. The guess parameters chosen are
$\left(A,~\Omega,~\phi\right)=\left(\Delta,~\Delta/10,~\pi/3\right)$
and those computed read $\left(A_{\rm opt},~\Omega_{\rm opt},~\phi_{\rm opt}\right)=
(5.838~\Delta,~0.211~\Delta,~28.173~{\rm rd})$.
$\Delta$ is an arbitrary unit of frequency.}
\label{fig:fig1}
\end{figure}
\section{Driven spin boson model}
The Hamiltonian of the driven spin boson model where
the two-level system is bilinearly coupled to an ensemble
of harmonic oscillators is given by~\cite{Legget,Grifoni}
\be
H=-\frac{\hbar}{2}\Delta\sigma_x -\frac{\hbar}{2}\left(\varepsilon_0+\varepsilon(t)\right)\sigma_z
+\frac{1}{2}\sum_{i}\left(\frac{p_i^2}{m_i} +m_i\omega_i^2 x_i^2\right)
+(\sigma_z\,q_0/2)\sum_{i}c_i\,x_i\,.
\ee
where $\sigma _{\alpha}$ with $\alpha=x,y,z$ are Pauli spin matrices; $\hbar\Delta $ 
is the tunneling splitting, $\hbar\varepsilon_0$ is an energy bias
and $\hbar\epsilon (t)$ is its modulation by a time-dependent external control
field; and the heat bath is represented by a set of harmonic oscillators of mass 
$m_{i}$, angular frequency $\omega _{i}$, momentum $p_{i}$
and position coordinate $x_{i}$. The oscillators are coupled
independently to the spin coordinate with strength measured by the set 
$\{c_{i}\}$ while $q_{0}$ measures the distance between the left and
right potential wells. 
The coupling constants enter in the spectral
density function of the environment defined by,
$
\label{eq:discrete_J_omega}
J(\omega)=\frac{\pi}{2}\sum_{i}\,{c_i\over m_i\,\omega_i}\,\delta(\omega-\omega_i)\, .
$

In order to compute the Redfield tensor, it is necessary
to determine the propagator of the coherent system dynamics
$U_S(t,\tau)$.
An analytical expression for $U_S(t,\tau)$ is not trivial
because the Hamiltonian of the physical system 
$H_S(t)=-\frac{\hbar}{2}\Delta\sigma_x -\frac{\hbar}{2}
\left(\varepsilon_0+\varepsilon(t)\right)\sigma_z$ is time-dependent and not diagonal.
To get round this difficulty, we transform the Hamiltonian
$H$ by the unitary operator, {\it i.e.}, polaron transformation
$
U= e^{\mathcal{O}}$ with $\mathcal{O}=-\frac{i}{2}\sigma_z\Omega$
and 
$\Omega =\sum_i\Omega_i$ where $\Omega_i= 
\left(q_0c_i/\hbar m_i\omega_i^2\right)p_i
$~\cite{Legget}. The transformed Hamiltonian $H'=UHU^{-1}$ takes the following form
\be
H'=-\frac{\hbar}{2}\left(\varepsilon_0+\varepsilon(t)\right)\sigma_z
+ \frac{1}{2}\sum_{i}\left(\frac{p_i^2}{m_i} +m_i\omega_i^2 x_i^2\right)
-\frac{1}{2}\hbar\Delta\left(\sigma_+ e^{i\Omega}+\sigma_- e^{-i\Omega}\right)\, ,
\ee
where $\sigma_{\pm}=(\sigma_x\pm i\sigma_y)/2$. After invoking 
the polaron transformation the coherent propagator is trivial and reads 
$U_S(t,t')= \cos\left[\varepsilon_0(t-t')/2+f(t,t')/2\right]{\mathbb{I}}
+ i\sin\left[\varepsilon_0(t-t')/2+f(t,t')/2\right]\sigma_z$
where the function $f(t,t')=\int _{t'}^{t}\,d\tau\varepsilon(\tau)$ 
captures the effects of the external control field.

\section{Rate master equation}
Let us consider the polaron transformed spin-boson Hamiltonian 
and derive the explicit form for the corresponding master equation for small $\Delta$.
Here, the system and the environment 
operators are $S_1=\hbar\Delta\sigma_+/2$, 
$S_2=S_1^{\dagger}=\hbar\Delta\sigma_-/2$ and $B_1=e^{-i\Omega}$,  
$B_2=B_1^{\dagger}=e^{i\Omega}$, respectively. 
The Bloch-Redfield formalism leads to the following master equation for the diagonal elements of the
the reduced density matrix (the populations)
\bea
\dot\rho_{11}(t) &=& \rho_{22}(t)W_{12}(t)- \rho_{11}(t)W_{21}(t)\,,\\
\dot\rho_{22}(t) &=& \rho_{11}(t)W_{21}(t)- \rho_{22}(t)W_{12}(t)\,,
\eea
with time dependent transition rates  
$W_{12}(t)=\Gamma_{21,12}^+(t) +\Gamma_{21,12}^-(t)$
and $W_{21}(t)= \Gamma_{12,21}^+(t) +\Gamma_{12,21}^-(t)$.
Up to the second order in $\Delta$, the populations decouple from the
non-diagonal terms. The master equation for the population difference
$z(t)=\langle\sigma_z\rangle_t=\rho_{11}(t)-\rho_{22}(t)$ reads
\be
\dot z(t) =-\Gamma_{zz}(t)z(t)+\Gamma_0(t)\,,
\ee
where $\Gamma_{zz}(t) = W_{12}(t)-W_{21}(t)$ and $\Gamma_0(t) = W_{12}(t)+W_{21}(t)$
defined as
\bea
\Gamma_{zz}(t)&=&\Delta^2\int_0^t\,d\tau\,e^{-Q_2(\tau)}\cos[f(t,t-\tau)]\cos[Q_1(\tau)]\,,\\
\Gamma_0(t)&=&\Delta^2\int_0^t\,d\tau\,e^{-Q_2(\tau)}\sin[f(t,t-\tau)]\sin[Q_1(\tau)]\,.
\eea
%
\begin{figure}
\onefigure[width=6cm,angle=-90]{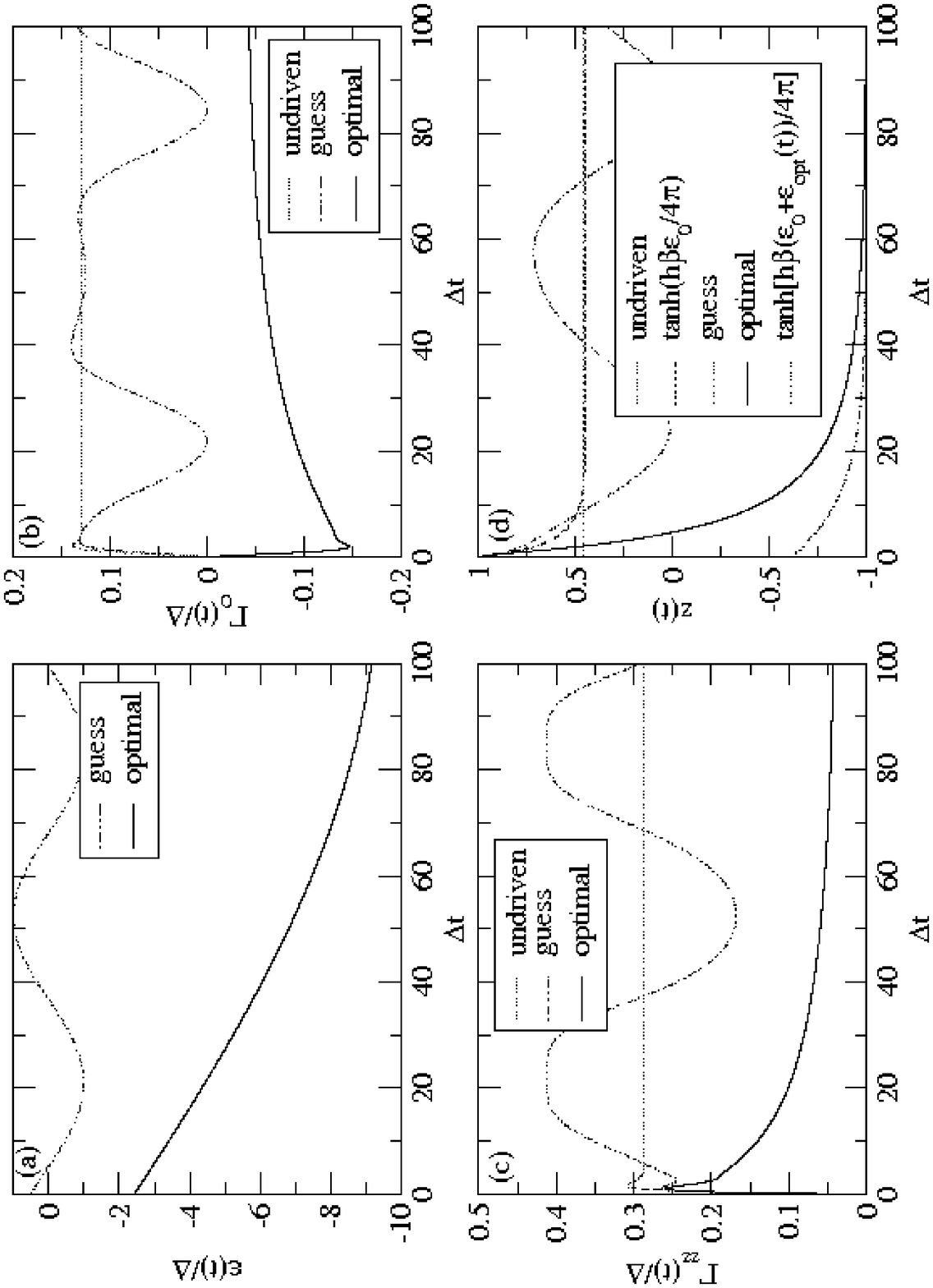}
\caption{Population transfer from the ground state 
$z(0)=1$ to the target excited $z(t_f)=-1$.
(a) shows the control field vs. time.
(b) and (c), respectively, show  $\Gamma_{zz}$ and $\Gamma_{0}$ vs. time.
(d) shows the relative population $z=\left\langle\sigma_z\right\rangle$ 
vs. time. $w_f=1$, $w_r=0$, $\alpha=0.2$, $\epsilon_0=\Delta$,
$\omega_c=20\Delta$, $\beta=1/\hbar\Delta$ 
and $t_f= 100/\Delta$. 
The guess parameters chosen are
$\left(A,\Omega, \phi\right)=\left(\Delta,\Delta/10,\pi/3\right)$
and those computed read $\left(A_{\rm opt},\Omega_{\rm opt},\phi_{\rm opt}\right)=
\left(9.34\Delta,-1.107\times{10^{-2}}\Delta, 10.73\,{\rm rd}\right)$.
$\Delta$ is an arbitrary unit of frequency.}
\label{fig:fig2}
\end{figure}
The quantities $Q_1(\tau)$ and $Q_2(\tau)$ are the imaginary and the real part, respectively,
of the function $\Phi(\tau)=\frac{q_0^2}{\pi\hbar}\int_0^{\infty}\,d\omega \frac{J(\omega)}{\omega^2}
\left\lbrack(1-\cos\omega\tau)\coth(\hbar\omega\beta/2)+i\sin\omega\tau)\right\rbrack$
where $e^{-\Phi(\tau)}=\langle e^{i\Omega(\tau)}e^{-i\Omega(0)}\rangle$
is the environment correlation function~\cite{Mahan}. 
In the present work, the spectral density of the 
environment $J(\omega)=\eta\omega e^{-\omega/\omega_c}$
is assumed to be Ohmic with exponential cutoff.
Here $\eta$ is a phenomenological friction coefficient.
For an Ohmic bath and at low temperatures regime ($\hbar\omega_c\beta \gg 1$)
one can use the approximation~\cite{Legget}
\bea
Q_1(\tau)&=& 2\alpha\arctan(\omega_c\tau),\\
Q_2(\tau)&=& \alpha\ln(1+\omega_c^2\tau^2)+
2\alpha\ln\left\lbrack\frac{\sinh(\pi\tau/\beta\hbar)}{(\pi\tau/\beta\hbar)}\right\rbrack\,,
\eea
where the dimensionless dissipation constant $\alpha=q_0^2\eta/2\pi\hbar$
have been introduced.
\section{Quantum optimal control problem}
Let time $t$ be in the interval $[0,t_f]$ for fixed $t_f$.
The evolution of the state variable $z(t)$ 
governed by the the master equation 
depends not only on the initial state $z(0)=z_i$
but also on the time-dependent control variable $\varepsilon(t)$.
The task is now to find a control field that will steer the system
from its initial state to a desired final state at specified time $t_f$.
Typically, it is possible to define a cost functional incorporating the
objective. The goal of optimal control algorithms is to
calculate a control field which can induce a specified system dynamics by 
minimizing this cost functional. Consider then the 
following quantum optimal control problem of minimizing the cost functional
$J({\bf p})$ subject to the dynamical constraint, {\it i.e}, 
the master equation for $z(t)=\langle\sigma_z\rangle_t$
\be
\begin{cases} 
{\rm min }\, J({\bf p}) = \frac{w_f}{2}(z(t_f)-z_d)^2 + \frac{w_r}{2t_f}\int_0^{t_f}\,dt\,(z(t)-z_r(t))^2\,,\\
\dot z(t) = -\Gamma_{zz}(t) z + \Gamma_0(t),\,\quad t \in [0,t_f],\\
z(0)=z_i\, .
\end{cases}
\ee
The first term in the cost functional $J(\bf p)$ measures the deviation
of the final state $z(t_f)$ from the desired state $z_d$ . 
During the time interval $[0,t_f]$, there may also exist a desired state
trajectory $z_r(t)$. This objective is incorporated in the second term. Here,  
it is assumed that $0\le w_f\le 1$ and $0\le w_r\le 1$
with $w_f+w_r=1$. $J({\bf p})$ is the sum of the so-called final 
time cost functional and running cost functional.
The vector ${\bf p}\in {\mathbb{R}}^{N_p}$ is a set of parameters on which the control field depends,
{\it i.e.}, $\varepsilon(t)\equiv\varepsilon(t,{\bf p})$. 
Here, an optimal solution of this problem is characterised by first order
optimality conditions in the form of the Pontryagin's Minimum
Principle~\cite{ARTHURE}. These conditions
are formulated with the the help of a Hamilton function
that has the following form in our problem:
\be
{\mathcal{H}}(z,\lambda,{\bf p})=\frac{w_r}{2t_f}{\left(z(t)-z_r(t)\right)}^2+\lambda(t)
\left\{-\Gamma_{zz}(t) z+\Gamma_0(t)\right\}\, .
\ee
Pontryagin's minimum principle states that a necessary condition 
for $(z,{\bf p})$ to be a solution of the above optimal control problem 
is the existence of an adjoint state $\lambda$ such that
\be
\begin{cases}
\dot z(t) = \frac{\partial \mathcal{H}}{\partial\lambda}=-\Gamma_{zz}(t) z + \Gamma_0(t)\,,\\
\dot\lambda(t)=-\frac{\partial \mathcal{H}}{\partial z}=-\frac{w_f}{t_f}(z(t_f)-z_d) +\Gamma_{zz}(t)\lambda(t)\,,\\
z(0)=z_i,\quad\lambda(t_f)= w_r\left(z(t_f)-z_d\right)\,,\\
0=\frac{\partial{\mathcal{H}}}{\partial p_i},\quad i=1\ldots N_p\quad\mbox{and}\quad t\in[0,t_f]\,.
\end{cases}
\ee
The minimum principle requires the solution of a set of complicated nonlinear algebraic equations,                                                
namely the optimality conditions ${\partial{\mathcal{H}}}/{\partial p_i}=0,\,i=1\ldots N_p$
which can only be solved in an iterative manner.
The present optimal control problem is not singular because 
$\det\left(\frac{\partial^2{\mathcal{H}}}{\partial p_i\partial p_j}\right)\not= 0,\, 
i,j=1\ldots N_p$, since $\Gamma_{zz}(t)$ and $\Gamma_{0}(t)$ 
depend nonlinearly on $\varepsilon(t,{\bf p})$. 
The optimality conditions ${\partial{\mathcal{H}}}/{\partial p_i }=0,\,i=1\ldots N$,
gives the gradient for the cost functional $J({\bf p})$ with respect 
to control parameters $p_i$
\be
\frac{\partial J}{\partial p_i}=\int_0^{t_f}\,
\lambda(t)
\left\lbrack
-\frac{\partial\Gamma_{zz}(t)}{\partial p_i}z(t)
+\frac{\partial\Gamma_0(t)}{\partial p_i}
\right\rbrack\,dt,\quad i=1\ldots N_p\,.
\ee
In summary, the computation of the gradient towards an optimal 
solution requires the computation of 
$z(t)$ forward in time and the adjoint state $\lambda(t)$
backward in time.  
\begin{figure}
\onefigure[width=6cm,angle=-90]{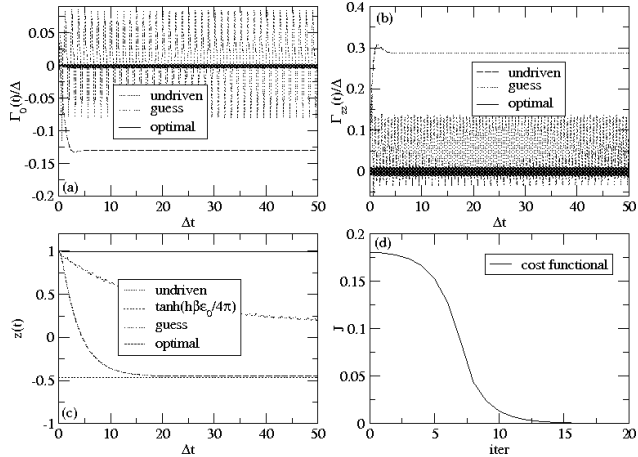}
\caption{Trapping of the spin in unstable excited state, $z(0)=1$.
(a) and (b), respectively, show  $\Gamma_{zz}$ and $\Gamma_{0}$ vs. time.
(c) shows the relative population $z=\left\langle\sigma_z\right\rangle$ vs. time.
(d) shows the cost functional vs. the number of iteration.
$w_f=0$, $w_r=1$, $\alpha=0.2$, $\epsilon_0=-\Delta$,
$\omega_c=20\Delta$, $\beta=1/\hbar\Delta$ 
and $t_f= 50/\Delta$. The guess parameters chosen are
$\left(A,\Omega,\phi\right)=\left(12\Delta,5\Delta,0\right)$
and those computed read $\left(A_{\rm opt},\Omega_{\rm opt},\phi_{\rm opt}\right)=
\left(218.196\Delta,~89.995\Delta,-1.117~{\rm rd}\right)$.
$\Delta$ is an arbitrary unit of frequency.
Note that $\Omega_{\rm opt}$ is much higher than $\omega_c$.
}
\label{fig:fig3}
\end{figure}
\section{Numerical Results}
In the present work, we write the control field as monochromatic plane wave 
$\varepsilon(t,{\bf p})\equiv\varepsilon(t,A,\Omega,\phi)= A\cos(\Omega t+\phi)$.
In this case, an analytic expression can be found for the function 
$f(t,t')\equiv f(t,t',A,\Omega,\phi)=\int _{t'}^{t}\,d\tau\varepsilon(\tau,A,\Omega,\phi)$
which greatly reduces the numerical effort for finding a solutions 
to this optimization problem.  
The task is then to find a set of three parameters namely
the amplitude $p_1=A$, the frequency ${p_2=\Omega}$ and the phase 
${p_3=\phi}$ such that the cost functional $J({\bf p})$ defined above is minimal.
The optimization is performed using the gradient method. More precisely, we used
the subroutine DMNG of port library~\cite{port} implementing the quasi-Newton method 
as variant of gradient algorithms.

\section{Heating}
As a first example, we consider driving the system from a pure state $z(0)=1$, 
corresponding to the temperature--zero ground state of the two--level system, into  
the mixed target state $z(t_f)=0$.
Fig. 1(d) shows that in the absence of the control, 
the system inevitably relaxes to the thermal equilibrium state 
\be
z_{\rm st}= \frac{\Gamma_{0}(\epsilon_0)}{\Gamma_{zz}(\epsilon_0)}
=\frac{W_{12}(\epsilon_0)-W_{21}(\epsilon_0)}{W_{12}(\epsilon_0)+W_{21}(\epsilon_0)}= 
\tanh(\hbar\beta\varepsilon_0/2)\, .
\ee
Fig. 1 shows also guess and optimal harmonic control, obtained via the gradient method,
the rate $\Gamma_{zz}(t)$ and the inhomogeneous term $\Gamma_0(t)$, 
as well as the time evolution of the relative population $z(t)$.  
Owing to the control dependence of $\Gamma_{zz}(t)$ and $\Gamma_0(t)$ 
which reflects the periodic nature of the control 
field, Fig.~1(b) and 1(c), the objective posed can be achieved perfectly. 
 
\section{Cooling}
As a second example, we consider a spin-flip in which 
$\left\langle\sigma_z\right\rangle =1$ at t=0 is changed to
$\left\langle\sigma_z\right\rangle=-1$ at target time $t_f$ 
with an optimal solution displayed in Fig.~2a.   
Here, the simplest optimum harmonic solution selected is a 
low-frequency field which simply changes the level sequence in the two-level system, 
at the same time, renormalising $\Gamma_{zz}$ and  $\Gamma_{0}$.  
Approaching target time $t_f$, 
$\Gamma_{zz}(t)$ approaches $-\Gamma_{0}(t)$ to provide $z(t_f)=-1$, 
as shown in Figs.~2(b) and 2(c). 
In Fig.~2(d) we also show  the instantaneous equilibrium value $z_{\rm st}(t)=
\tanh\left\lbrack\hbar\beta\left(\varepsilon_0+\varepsilon_{\rm opt}(t)\right)/2\right\rbrack$
which corresponds to the ratio 
$\left(W_{12}(t)-W_{21}(t)\right)/\left(W_{12}(t)+W_{21}(t)\right)$\break
$=\Gamma_0(t)/\Gamma_{zz}(t)$
relevant in the adiabatic limit, valid when $\left|d\varepsilon(t)/dt\right|\ll\Delta^2$
as discussed in the literature~\cite{Luban,Grifoni}. 
In the long-ime limit, the prediction from the adiabatic approximation is fulfilled by 
the numerically selected optimal electric field $\varepsilon_{opt}(t)$  
in Fig.~2(d). 
In the context of bath-assisted cooling of spins,
our finding has certain analogies with the results of Ref.~\cite{Allahverdyan}
where a version of the spin-boson model under influence of externally controlled pulses is studied. 
Similar to our situation, an interference between the external fields 
and the spin-bath interaction create a mechanism that cool the spins much below the 
bath temperature. Starting from the initially infinite-temperature spin 
the authors cool it down to very low temperatures by increasing the spin's polarisation.

\section{Trapping }
As a third example, we consider trapping of the spin system for 
$\varepsilon_0=-\Delta$ in state $z(t)=1$ for all time $t \in [0,t_f]$ which now 
corresponds to the (unstable) excited  state of the system.
In isolated case, trapping is made possible by dynamic localization using a periodic 
control field to renormalize the coupling between the two levels
from $\Delta$ to $\Delta J_0(A/\Omega)$ for a large frequency of the field;
here $J_0(x)$ is the zero order Bessel function~\cite{Grifoni,Stockburger}.
Therefore, when $A/\Omega$ is such that $J_0(A/\Omega)=0$, the two-levels are decoupled
and the tunneling between them is suppressed.
In the presence of dissipation the trapping is destroyed, 
the undriven system relaxes to the equilibrium state, 
and the guess harmonic field with high frequency is not 
able to trap the system as shown in Fig.~3(c).  
For an optimum harmonic control field, we find 
that trapping becomes truly effective when the control frequency exceeds 
the cut--off frequency which was set as $\omega_c=20\Delta$.
The optimum solution for the control  parameters is 
$\left(A_{\rm opt}, \Omega_{\rm opt}, \phi_{\rm opt}\right)=
\left(218.196~\Delta, 89.996~\Delta,-1.119~{\rm rad}\right)$.
As shown in Figs.~3(a) and 3(b), this high-frequency control renormalizes 
$\Gamma_{zz}(t)$ and $\Gamma_{0}(t)$ 
to values of the order $10^{-5}\Delta$ such that $\dot z(t)\approx 0$
for all time $t \in [0,t_f]$.  
This finding is in qualitative agreement with the bang--bang method  
used to  study the independent spin-boson model~\cite{Viola}.
As we mentioned in the introduction, this study showed that the control has to be switched 
at a rate greater than $1/\omega_c$ in order 
to provide effective decoupling of the physical system
from the environment. Fig.~3(d) shows the cost functional versus 
the number of iteration. According to this figure, 
the cost functional is monotonically 
decreasing and its convergence to zero is reached after $15$ iterations.

\section{Conclusion}
In summary, we have formulated and solved an optimal 
control problem for the spin--Boson model
to demonstrate feasibility of controlling the effective 
system--bath interaction by an external control field.
An external control not only renormalises the spin Hamiltonian 
but also the effective coupling between system and environment.    
We have demonstrated that the effective system-bath interaction 
can be increased or decreased in controlled fashion. 
This mechanism  can be used to optimize  
dynamic driving and trapping of the spin system.  
Results were presented for control of the relative 
population $z$ of the spin system and a 
harmonic control field.  
This work has been extended to a control of 
all components of the Bloch vector simultaneously, 
as well as to general control field shapes. 
The results will be presented elsewhere.

\stars

We wish to acknowledge financial support of this work by FWF, project 
number P16317-N08.

\end{document}